\documentclass[12pt,a4paper]{article}
  
\usepackage{amsmath,amssymb,graphicx,float,colordvi,url,wrapfig,color}

\def\ta{\textcolor{black}}

\begin{document}

\thispagestyle{empty}
\addtocounter{page}{-1} 
\vskip-5cm 
\vspace*{0.6cm} 
\centerline{\Large \bf Unruh effect in a real scalar field with the \textcolor{black}{Higgs type} potential}
\vspace*{0.4cm}
\centerline{\Large \bf \textcolor{black}{on the  de Sitter space}}

\vspace*{1.0 cm}  
\centerline{\large \bf Shingo~Takeuchi}
\vspace*{0.7 cm}
\centerline{\it The Institute for Fundamental Study, ``The Tah Poe Academia Institute''}
\vspace*{0.15 cm}
\centerline{\it Naresuan University Phitsanulok 65000, Thailand}
\vspace*{1.0 cm}
\centerline{\bf Abstract} 
\vspace*{0.4 cm} 
It \textcolor{black}{has been} predicted~\textcolor{black}{\cite{Experiment,Sciama,Raval:1995mb,Chen:1998kp,Iso:2010yq}} that 
an accelerating electron performs a Brownian motion in the inertial frame. This Brownian motion in the inertial frame 
has its roots in the interaction with the thermal excitation given by the Unruh effect in the accelerating frame. 
If such a prediction is possible, we correspondingly propose a prediction in this study that the thermal radiation appears 
in the inertial frame from an electron heated by the Unruh effect in the accelerating frame. The point in our prediction is, 
although the Unruh effect is \textcolor{black}{only} in the accelerating frame, 
\textcolor{black}{if the appearance of the Brownian motion rooted in the Unruh effect in the inertial frame can be predicted,  
the heat that the particle gets in its body by the Unruh effect in the accelerating frame could survive in the inertial frame.} 
Based on such a prediction, in this paper we investigate phenomena in the neighborhood of an accelerating electron in the inertial frame. 
The model we consider is the four-dimensional Klein-Gordon real scalar field model with the \textcolor{black}{Higgs type} potential term at the finite temperature 
identified with the Unruh temperature on the de Sitter space-time. We calculate the one-loop effective potential in the inertial frame 
with the corrections by the thermal radiation rooted in the Unruh effect in the accelerating frame. In this calculation, 
we take into account that the background space-time is deformed due to the field theory's corrected one-loop effective potential. 
Based on such an analysis, we illustrate the restoration of the spontaneous symmetry breaking and the variation of the background space-time, 
and we examine the accelerating particle's world-line and the amount of the energy corresponding to the change of the acceleration.

\newpage
 
\section{Introduction}
\label{Chap:Intro}  

\textcolor{black}{The} ultra high intensity lasers have been developed significantly in these days, which are said to enable the experimental confirmations 
for the theoretically predicted quantum effects~\cite{Experiment}. One of them is the Unruh effect~\cite{Davies:1974th,Unruh:1976db,Unruh:1983ac}. 
It is a prediction that one moving in the Minkowski space-time with linear constant acceleration experiences the space-time as a thermal
bath given by a canonical ensemble with the Unruh temperature, $T_U = \hbar\, a/(2\pi c \,k_B) \approx 4 \times 10^{-23}\,a\,/\,({\rm {1}\,cm}/s^2)\,[{\rm K}]$, 
where $a$ is the acceleration.

In the above situation, we think that it is timely and meaningful to investigate the theoretical aspect of the phenomena of the Unruh effect. 
There are already considerations and attempts to detect the Unruh effect~\cite{Experiment,Sciama,Raval:1995mb,Chen:1998kp,Iso:2010yq}. 
These exploit the Larmor radiation that an accelerating electron emits. Here the Larmor radiation they exploit is an additional Larmor
radiation emitted by the Brownian motion rooted in the thermal bath given by the Unruh effect in the accelerating frame,
in addition to the normal Larmor radiation. Such an additional Larmor radiation is called ``{\it Unruh radiation}'' in these papers.

One of the interesting points \textcolor{black}{in} the Unruh radiation is that, 
although the Unruh effect itself is a phenomenon occurring only in the accelerating frame, 
the Brownian motion rooted in the thermal bath given by the Unruh effect in the accelerating frame is predicted in the inertial frame, 
and as a result the Unruh radiation is emitted in the inertial frame. 
\textcolor{black}{If such an additional Larmor radiation~(the Unruh radiation) could be predicted in the inertial frame, 
it could be also predicted that the heat that the accelerating particle gets in the accelerating frame by the Unruh effect can survive in the inertial frame,  
and there appears the thermal radiation from the heat of the accelerating particle in the inertial frame}.

In this study, we hence propose a prediction mentioned above. The point in our prediction is, although the Unruh effect occurs only in the accelerating frame, 
\textcolor{black}{corresponding to the appearance of the Unruh radiation  in the inertial frame that is rooted in the Unruh effect, 
the heat taken in the body of the particle by the Unruh effect in the accelerating frame could survive if the frame is changed to the inertial frame, 
and as a result the thermal radiation rooted in the Unruh effect could appear from the accelerating particle even in the inertial frame}. 
Namely, although the Unruh effect itself is the specific only in the accelerating frame, the resultant heat in the body of the particle in the accelerating frame 
could survive even in the inertial frame. We call such a thermal radiation ``{\it Unruh thermal radiation}''.

Let us here turn to the recent detection of the Higgs particle~\cite{Aad:2012tfa,Chatrchyan:2012ufa}. 
This leads us to the prediction that the realistic field's vacuum energies are given by some Higgs potentials.
Since the spontaneous symmetry breaking described by the Higgs potentials is restored by the thermal effect~\cite{Dolan:1973qd}, 
the Unruh thermal radiation can also restore the spontaneous symmetry breaking. In this case the vacuum energies with such a thermal correction affect 
the structure of the space-time. Because it can play the role of the cosmological constant.
\newline

Hence, the purpose of this study is, based on our prediction for the Unruh thermal radiation mentioned in the fourth paragraph, 
\textcolor{black}{to calculate the Higgs type field theory's vacuum energy in the neighborhood of the accelerating particle with the thermal correction 
given by the the Unruh thermal radiation that is emitted from the particle, and becomes the cause for the restoration of the spontaneous symmetry breaking.} 
In this calculation we take into account the deformation of the background space-time to the de Sitter space due to the field theory's corrected vacuum energy. 
Based on such an analysis, we investigate \textcolor{black}{some related phenomena}.

We now \textcolor{black}{mention more specifically} what we will do. We first consider a particle moving with a linear constant acceleration 
in the four-dimensional Minkowski space-time. \textcolor{black}{Then, if the prediction for the Brownian motion rooted in the Unruh effect 
is possible in the inertial frame, we propose a prediction in this paper that there is the thermal radiation from the accelerating particle 
in the inertial frame, which is rooted in the Unruh effect, and we call it ``{\it Unruh thermal radiation}'' as in the fourth paragraph}. 
As a result, in the inertial frame, the field in the neighborhood of the particle gets heated and the field's vacuum energy gets a thermal correction. 
The field we consider in this study is a real scalar field described by the Klein-Gordon equation with a \textcolor{black}{Higgs type} potential term, 
at finite temperature, where the temperature is brought \textcolor{black}{into} according to the imaginary time formalism, and it is identified with the Unruh temperature. 
We then calculate the one-loop effective potential with the correction given by the Unruh thermal radiation, 
where the spontaneous symmetry breaking in the corrected one-loop effective potential is restored finally by the Unruh thermal radiation's correction. 
We perform this calculation \textcolor{black}{with} taking into account that the background space-time is deformed to the de Sitter space 
due to the corrected one-loop effective potential \textcolor{black}{working} as the cosmological constant. Although the cosmological constant in our actual cosmology is not zero, 
taking the approximation that the space-time will be flat at zero temperature for simplicity in \textcolor{black}{our analysis}, 
we define the corrected one-loop effective potential to vanish at zero temperature. 
Based on such a corrected one-loop effective potential, we examine the amount of the energy corresponding to the change of the acceleration 
and the accelerating particle's world-line for each acceleration. 
\newline

We refer to a few basic papers on the spontaneous symmetry breaking and its restoration induced by the thermal effect and Unruh effect. 
In Ref.\cite{Dolan:1973qd}, the symmetry breaking and restoration in a flat space at finite temperature have been investigated. 
In Ref.\cite{Castorina:2012yg}, the restoration of the spontaneous symmetry breaking in a real scalar field in a Rindler space has been investigated. 
In Refs.\cite{Ohsaku:2004rv} and \cite{Ebert:2006bh}, 
the chiral condensation and the quark and diquark condensation in the Nambu-Jona-Lasinio model in the Rindler space have been investigated, respectively. 
\textcolor{black}{In Ref.\cite{Takeuchi:2015nga}, the dissolution of the Bose-Einstein condensate in the free complex scalar field model at finite density 
in a Rindler space has been investigated.} There are also papers concluding that the Unruh effect does not restore the spontaneous symmetry breaking in 
the Rindler space~\cite{Hill:1985wi}.

One may notice that the symmetry restorations in Refs.\cite{Castorina:2012yg,Ohsaku:2004rv,Ebert:2006bh,Takeuchi:2015nga} are the ones in the accelerating frame. 
On the other hand, the symmetry restoration in this paper is the one in the inertial frame. Further, the phenomena in the situation that
the spontaneous symmetry breaking is being restored with the deformed background space-time due to the corrected field's vacuum energy by the Unruh 
thermal radiation have never been investigated so far, and we first investigate these in this paper. As for Ref.\cite{Hill:1985wi}, 
their conclusion does not apply to this study readily. Because their issue is how the restoration of the spontaneous symmetry breaking is in the
accelerating frame. On the other hand, the issue we address in this paper is a phenomenology in the inertial frame provided that a particle becomes 
hot in the inertial frame by the Unruh effect in the accelerating frame. Further, the thermal source in the restoration of the spontaneous symmetry 
in this paper is not the Unruh effect itself but the heat of the particle in the inertial frame rooted in the Unruh effect. As long as they can provide 
some mechanism that can cancel the Unruh effect in the accelerating frame, their conclusion does not apply to our study readily. 
\newline

For the reason that the effective potentials in the realistic fields can be predicted to be some Higgs potential type, there is a certain probability that 
the phenomena similar to what we investigate in this paper are actually occurring in our world. \textcolor{black}{Therefore, this study would be important and intriguing}.
\newline

\textit{Note}~:~After this paper has been published, the author found the papers that address the issue on a thermal property~(the energy equipartition relation) 
of an accelerated particle and the radiation from the accelerating particle with the thermal Brownian motions~\cite{Oshita:2014dha,Oshita:2015xaa}. 
These are extended to the case in the particle coupling to the electromagnetic field as more realistic situation~\cite{Oshita:2015qka}. 
The authors in Refs.\cite{Oshita:2014dha,Oshita:2015xaa,Oshita:2015qka} claim that the naively expected thermal radiation does not exist 
and the quantum radiation rooted in the Unruh effect of the accelerating frame could survive~(the quantum radiation mentioned here means 
the Unruh radiation first given in the second paragraph). Here, the quantum radiation and the Larmor radiation from the classical accelerated 
motion \textcolor{black}{are partially canceled} each other, and the quantum radiation finally appears with being suppressed for that, 
according to their personal comment. Their stuffs would be related with the Unruh thermal radiation in the inertial frame that we predict in this paper.
\newline



We mention the organization of this paper. In Sect.\ref{Chap:The model}, we give the background space-time and the model in the field theory
in this study. In Sect.\ref{Chap:One-loop effective potential}, introducing the finite temperature according to the imaginary formalism, 
we calculate the one-loop effective potential in the field theory at the finite temperature, and then we define it such that it can vanish at zero
temperature. In Sect.\ref{Chap:Fixing the background space-time}, by regarding the one-loop effective potential obtained in Sect.\ref{Chap:One-loop effective potential} 
\textcolor{black}{as the cosmological constant}, we give the relation to fix the background space-time. 
In Sect.\ref{Chap:Result}, we show the results in this study: the Hubble constant that characterizes the variation of our de Sitter space, 
and the variation of the shape of the one-loop effective potential, against the temperature, 
and after that the amount of the energy corresponding to the change of the temperature at each temperature

\section{The model}
\label{Chap:The model} 

We start with the following action:
\begin{eqnarray}\label{scalar action}
S^M &=& \int \! d^4 x \sqrt{-g^M} \, 
\left( 
R^M + \frac{1}{2} \partial_\mu \phi \, \partial^\mu \phi \,-\, \frac{1}{2} m^2 \phi^2 \,-\, \frac{\lambda}{4\,!}\phi^4
\right)\nonumber\\
&\equiv& 
\int \! d^4 x \sqrt{-g^M} \, \left(R^M + {\cal L}^M \right),
\end{eqnarray}
where $R$ and $\phi$ are the scalar curvature and the real scalar field, respectively. 
$m^2$ is some \textcolor{black}{negative} constant in order to have a \textcolor{black}{Higgs type} potential. 
In this paper, we attach ``$M$'' to the quantities given in the Minkowski space-time. 
On the other hand, we denote the quantities given in the Euclid space without ``$E$'', 
except for the case that we need to distinguish it explicitly.

It will turn out later that the field's effective potential plays the role of a positive cosmological constant. 
Hence we take the four-dimensional de Sitter space-time,
\footnote{
We consider the finite temperature according to the imaginary time formalism. Thus, the de Sitter space in \textcolor{black}{the} static coordinates, 
$
ds_M^2 = \left( 1 - {r^2}/{\alpha^2} \right) dt^2 - \left( 1 - {r^2}/{\alpha^2} \right)^{-1} dr^2 - r^2 \left( d\theta^2 + \sin^2 \theta \,d \phi^2 \right) 
$, 
is convenient, because the metric is time-independent. 
($\alpha$ is the radius of the de Sitter space in a Minkowski space-time, with one higher dimension, in which the de Sitter space-time is embedded.)  
However, for convenience in the actual analysis, we take a physically equivalent other coordinate, the flat slicing coordinate~(\ref{de sitter}). 
Even in this coordinate, we can get a time-independent effective potential as mentioned under Eq.(\ref{Omega0}). 
}
\begin{eqnarray}\label{de sitter}
ds_M^2 = dt^2 - e^{2H^M t}\left( dx^2 +dy^2+dz^2 \right),  
\end{eqnarray} 
as our background space-time, where the Hubble constant $H^M$ is not fixed at this stage.

We now address the case of a particle moving with a linear constant acceleration $a$. 
Following the Unruh effect's prediction, the particle experiences the space-time where the particle exists as a thermal bath with the Unruh temperature,
\begin{eqnarray}\label{TUnruh}
T_U = \frac{a}{2\pi},
\end{eqnarray} 
in the accelerating frame~(we use the natural units.). Following our prediction for the Unruh thermal radiation mentioned in the fourth paragraph of Sect.\ref{Chap:Intro}, 
the particle is thought to have some temperature rooted in the Unruh temperature, and to emit thermal radiation \textcolor{black}{in the inertial frame}. 
The thermal radiation \textcolor{black}{will} gradually decrease \textcolor{black}{as} getting away from the particle due to thermal dissipation in nature, 
and the radiation range \textcolor{black}{will be} finite. However in this paper, we take the radiation range from zero to infinity for simplicity in the analysis. 
We further regard the temperature distribution as uniform in the entire region.

\section{One-loop effective potential at finite temperature}
\label{Chap:One-loop effective potential} 

We calculate the one-loop effective potential of the real scalar field at finite temperature. To introduce the temperature, 
we perform a Wick rotation, $t \to -i\,\tau$, and \textcolor{black}{impose a} periodicity with the period $\beta = 1/T$ to the imaginary time direction. 
This $T$ is identified in this study with the Unruh temperature $T_U$ given in Eq.(\ref{TUnruh}) as mentioned in the last paragraph of Sect.\ref{Chap:The model}. 
Then the background geometry (\ref{de sitter}) is transformed as
\begin{eqnarray}\label{Euclid de sitter}
ds_M^2 \rightarrow ds^2 = -d\tau^2 -e^{2H\,\tau}\left( dx^2 +dy^2+dz^2 \right) 
\end{eqnarray}
with $H^M \to i H$, where we \textcolor{black}{considered} the Hubble constant \textcolor{black}{$H^M$} in the usual form \textcolor{black}{written by} a scale factor and its time derivative
\footnote{\textcolor{black}{Generally the usual notation of the Hubble constant $H^M$ using the scale factor $a$ is $H^M=\dot{a}/a$.}} 
~\textcolor{black}{(In this paper we refer both $H$ and $H^M$ as the Hubble constant with no distinction)}, and the symbol ``$\rightarrow$'' means the manipulation of the Wick rotation. We use it in this sense in this paper.  
Correspondingly the probability amplitude with the action (\ref{scalar action}) is changed to the partition function by
\begin{eqnarray}
Z
\!\!\!&=&\!\!\! 
\int {\cal D}\phi \, \exp \left( i \int \! d^4x \sqrt{-g^M} \left( R^M + {\cal L}^M \right) \right)\\
\!\!\!&\rightarrow&\!\!\! 
\int {\cal D}\phi \, \exp \left(\int_0^\beta \! d\tau \! \int \! d^3x \, \sqrt{g} \, \left( R + {\cal L} \right) \right)\nonumber\\
\!\!\!&=&\!\!\! 
\exp \left(\int_0^\beta \! d\tau \! \int \! d^3x \, \sqrt{g} \, R \right)
\cdot \int {\cal D}\phi \, \exp \left(\int_0^\beta \! d\tau \! \int \! d^3x \, \sqrt{g} \, {\cal L} \right)\nonumber\\
\!\!\!&\equiv&\!\!\! 
\exp \left( - \int_0^\beta \! d\tau \! \int \! d^3x \, \sqrt{g} \, \big(-R \big) \right) \cdot Z_\phi. 
\end{eqnarray}
We denote the determinant of the metric and the Lagrangian density after the transformation as $g$ and ${\cal L}$, respectively.
\newline

We now consider the quantum fluctuation of $\phi$ as $\phi= \phi_0 + \delta \phi$, where $\phi_0$ is the scalar field's condensation 
independent of the space-time at the symmetry breaking phase, and $\delta \phi$ is the quantum fluctuation. Then $Z_\phi$ can be written as
\begin{eqnarray}
Z_\phi \!\!\!&=&\!\!\!  \int {\cal D}\phi \, \exp \left( \int_0^\beta \! d\tau \! \int \! d^3x \, \sqrt{g} \, {\cal L} \right)\nonumber\\
&\!\!\!=\!\!\!& 
\exp \left( - \int_0^\beta \! d\tau \! \int \! d^3x \, \sqrt{g} \, {\cal L}_0 \right) \cdot
\int {\cal D}(\delta\phi) \, \exp \left( - \frac{1}{2} \int_0^\beta \! d\tau \! \int \! d^3x  \, \delta\phi \, G^{-1} \, \delta\phi \right),
\end{eqnarray}
where
\begin{eqnarray}
{\cal L}_0 = \frac{m^2}{2} \,\phi_0^2 + \frac{\lambda}{4!} \,\phi_0^4,
\end{eqnarray}
\vspace{-6mm}
\begin{eqnarray}\label{EOMOp}
G^{-1}  \!\!\!&=&\!\!\! e^{H \tau}\partial_z^2+\sqrt{g}\,(M^2 -3H\partial_\tau -\partial_\tau^2).   
\end{eqnarray}
Here
\begin{eqnarray}\label{CM2}
M^2 \equiv m^2+\frac{\lambda}{2}\,\phi_0^2.  
\end{eqnarray}
Then integrating out the $\delta \phi$, we obtain 
\begin{eqnarray}
Z_\phi
\!\!\!&=& \!\!\!  
\exp \left( - \int_0^\beta \! d\tau \! \int \! d^3x \, \sqrt{g} \,{\cal L}^{\rm E}_0 - \frac{1}{2} \, \log \,{\rm Det} \, G^{-1} \right)\nonumber\\
\!\!\!&\equiv& \!\!\!  
\exp(-\Gamma_E).
\end{eqnarray}

Performing the Fourier transformation and evaluating the functional determinant, we obtain
\begin{eqnarray} 
\Gamma_E\!\!\!&=&\!\!\!
\int_0^\beta \! d\tau \! \int \! d^3x \, \sqrt{g} \,\, {\cal L}_0 \nonumber\\
&&\!\!\!+\,\frac{1}{2\beta\,(2\pi)^3} 
\sum _{n=-\infty }^{\infty} 
\int \! d\tau \, d^3x \, d^3k \, \log 
\Big( e^{H\tau}k^2+\sqrt{g}\left( M^2-3iH\omega_n + \omega_n^2\right)\!\Big)\nonumber\\
\!\!\!&\equiv&\!\!\! \int \!d^3x \cdot V. \label{VolVE}
\end{eqnarray}
where $\omega_n \equiv 2\pi n/\beta$~($n$ is \ta{integer}). We make one comment. 
Now we have factored out $\int d^3x$. 
If factoring out the volume factor, the factor $\int d^4x \sqrt{g}$ should be factored out. 
However our integrand function depends on \textcolor{black}{$\tau$}. 
Hence factoring out of all volume cannot be done at this moment. 
We \textcolor{black}{will} however factor out the rest, $\int d\tau \sqrt{g}$, \textcolor{black}{under} Eq.(\ref{Omega0}).

We now evaluate the summation of $n$. To this aim, we rewrite the part concerned as
\begin{eqnarray} 
&& \sum _{n=-\infty }^{\infty} \log 
\Big(e^{H \tau} k^2  + \sqrt{g}\, \left( M^2 - 3i H \omega_n + \omega_n^2 \right)\!\Big)\nonumber\\
&\!\!\!=\!\!\!& 
\lim_{n \to \infty}3H \tau \, (2n+1) \,+\! \sum _{n=-\infty}^{\infty} \log \Big( e^{-2H \tau} k^2 + M^2 - 3i H \omega_n + \omega_n^2 \Big)\ta{.} 
\end{eqnarray} 
Then once performing the derivative as $\partial_\chi \log \left(\chi^2 - 3i H \omega_n + \omega_n^2 \right)$, where $\chi^2 \equiv e^{-2H \tau} k^2 + M^2$, 
we perform the summation. After that we integrate it. As a result, we obtain 
\begin{eqnarray}
V = V_{0}+V_{1},
\end{eqnarray}
where
\begin{eqnarray}
V_{0} \!\!\!&=&\!\!\! \int_0^\beta \! d \tau \, e^{3 H \tau } \, {{\cal L}_0},\\
V_{1} \!\!\!&=&\!\!\! \frac{1}{(2\pi)^2 \beta} 
\int \! d\tau \, dk \, \sqrt{g} \,\, k^2 \log \left[
-\sinh \left( \frac{\beta}{4} \left( \sqrt{9 H^2 + 4\left( e^{-2H \tau} k^2 + M^2 \right) }-3 H\right)\right) \right.
\nonumber\\
&& \qquad\qquad\qquad\qquad\quad\qquad
\left. \times \sinh \left( \frac{\beta}{4} \left( \sqrt{9 H^2 + 4\left( e^{-2H \tau} k^2 + M^2 \right) }+3 H \right)\right) \right] \nonumber\\
&& \qquad\qquad\qquad\qquad\qquad 
+ \frac{1}{2(2\pi)^3 \beta} \int \! d\tau \, dk^3 \lim_{n \to \infty}3H \tau\,(2n+1).  
\end{eqnarray}
Performing the rescaling $k \to e^{H \tau}\, k$, we can write $V_{1}$ as
\begin{eqnarray}
V_{1} \!\!\!&=&\!\!\! \frac{1}{(2\pi)^2 \beta} \int \! d\tau \, dk \, \sqrt{g} \,\, k^2 \, \log \left[
-\sinh \left(\frac{\beta}{4} \left(\sqrt{9 H^2+4 \left(k^2+M^2\right)}-3 H \right)\right)\right. \nonumber\\
&& \qquad\qquad\qquad\qquad\quad\qquad\,
\left. \times \sinh \left( \frac{\beta}{4} \left( \sqrt{9 H^2 +4 \left(k^2+M^2\right)}+3 H \right)\right) \right] \nonumber\\
&& \qquad\qquad\qquad\qquad\qquad 
+ \frac{1}{2\beta\,(2\pi)^3} \int \! d\tau \, dk^3 \lim_{n \to \infty}3H \tau\,(2n+1).  
\end{eqnarray} 
Pushing forward the calculation, we can write $V_{1}$ as 
\begin{eqnarray}
V_{1}=V_{1,\,(T=0)}+V_{1,\,(T>0)},
\end{eqnarray}
where $V_{1,\,(T=0)}$ and $V_{1,\,(T>0)}$ are the contributions independent of the temperature and depending on the finite temperature as
\begin{eqnarray}
\label{V_1(T=0)}
V_{1,\,(T=0)} \!\!\!&=&\!\!\! \frac{1}{4} \int_0^\beta \! d\tau \,\sqrt{g}\, \int \! \frac{d^3k}{(2\pi)^3}\,\sqrt{9 H^2 +4 \left(k^2+M^2\right)}\ta{,}\\
\label{V_1(T>0)}
V_{1,\,(T>0)} \!\!\!&=&\!\!\!
\frac{1}{(2\pi)^2\beta} \int_0^\beta \! d\tau \,\sqrt{g}\, \int \! dk \, 
\,k^2\,\log \left[
\left(1-e^{-\frac{\beta}{2} \left(\sqrt{9 H^2 +4 \left(k^2+M^2\right)}-3 H\right)}\right)\right. \nonumber\\
&& \qquad \qquad\qquad\qquad\qquad\quad\,\, \times \left.
\left(1-e^{-\frac{\beta}{2} \left(\sqrt{9 H^2 +4 \left(k^2+M^2\right)}+3 H\right)}\right)\right]+\Omega
\end{eqnarray}
and
\begin{eqnarray}\label{Omega0}
\Omega = \frac{1}{2\beta\,(2\pi)^3} \int \! d\tau \, dk^3 \left( \lim_{n\to \infty } 3 H \tau (2 n+1) + e^{3H \tau}(-2 \log (2)+ i \pi )  \right).
\end{eqnarray}

Now it can be seen that we can factor out \textcolor{black}{as} $V \rightarrow \int d\tau \, \sqrt{g} \cdot V$. 
By combining this with the factoring out of the space part performed in Eq.(\ref{VolVE}),  
we can now think \textcolor{black}{that the factoring out of the whole volume factor has been done}.

Let us next turn to the squared mass in the effective potential, which is the coefficient of $\phi_0^2\ta{/2}$ and currently given as $m^2$. 
It is assumed that it should be written as $m_{\rm eff}^2 = m^2+ \, \delta m^2$ after evaluating the finite temperature contribution, 
where $m_{\rm eff}^2$ is generally called the squared effective mass, 
and $\delta m^2$ formally represents the finite temperature contribution to the coefficient of $\phi_0^2{/2}$.~(Our result at the one-loop order is given in Eq.(\ref{Meff2}).)  
Using this $m_{\rm eff}^2$, $M^2$ given in Eq.(\ref{CM2}) is thought to be indeed given by $M_{\rm eff}^2 \equiv m_{\rm eff}^2+\frac{\lambda}{2}\phi_0^2$.  
Correspondingly, the $M^2$ in Eq.(\ref{V_1(T>0)}) \textcolor{black}{is} rewritten $M_{\rm eff}^2$. 
We expand $V_{1,\,(T>0)}$ around $M_{\rm eff}^2=0$  to the linear order. 
Around the symmetry breaking restoration, $m_{\rm eff}^2$ and $\phi_0^2$ take small values, and $M_{\rm eff}^2$ takes small values.

Finally, the one-loop effective potential we obtain with the factor out of the volume factor done above is
\begin{eqnarray}\label{EP_03}
V = V_{0}+V_{1,\,(T=0)}+V_{1,\,(T>0)},
\end{eqnarray}
where now 
$\displaystyle     
Z_\phi = \exp \left(-\int_0^\beta \! d\tau \! \int \! d^3x \, \sqrt{g} \cdot V \right)
$ 
and 
\begin{eqnarray}
\label{EP_03_tree}
V_{0} \!\!\!&=&\!\!\! {{\cal L}_0},\\
\label{EP_03_1LZT}
V_{1,\,(T=0)} \!\!\!&=&\!\!\! \frac{1}{4} \int \! \frac{d^3k}{(2\pi)^3}\,\sqrt{9  H^2 +4 \left(k^2+M_{\rm eff}^2\right)},\\
\label{EP_03_1LFT}
V_{1,\,(T>0)} \!\!\!&=&\!\!\!
\frac{1}{(2\pi)^2} \int dk \, k^2 \Bigg\{ \frac{1}{\beta} \,
\log \left[ 
\left(1-e^{-\frac{\beta}{2} \left(\sqrt{9 H^2+4 k^2}-3 H\right)} \right) 
\left(1-e^{-\frac{\beta}{2} \left(\sqrt{9 H^2+4 k^2}+3 H\right)} \right)
\right] \nonumber\\
&&
+ \,  
\frac{1}{\sqrt{9 H^2+4 k^2}}
\left.\left(
 \frac{1}{e^{\frac{\beta}{2}  \left(\sqrt{9 H^2+4 k^2}+3 H\right)}-1}
+\frac{1}{e^{\frac{\beta}{2}  \left(\sqrt{9 H^2+4 k^2}-3 H\right)}-1}
\right)M_{\rm eff}^2 \right\} + \Omega
\nonumber\\ 
\end{eqnarray}
and
\begin{eqnarray} \label{EP_03_1LFTC}
\Omega = \frac{1}{2\beta \,(2\pi)^3} \, \int\!dk^3 \left( \frac{1}{V_\tau} \int^\beta_0\!d\tau\,\lim_{n \to \infty} 3 H \tau (2n+1) -2 \log (2)+ i \pi \right).
\end{eqnarray}
Here, $V_\tau$ is defined as $\displaystyle V_\tau \equiv \int^\beta_0 d\tau \sqrt{g}=\left(-1+e^{H\,\beta}\right)/3H$. 
The above may agree with the case of flat space at $H=0$ 
\footnote{
Eq.(\ref{EP_03_1LFT}) can be regarded as the result in the high temperature expansion in the condition $|M_{\rm eff}/T| \ll 1$. 
Eq.(\ref{EP_03_1LFT}) at $H=0$ can be written as
\begin{eqnarray}\label{EP_HighTempExp_Flat}
\left. V_{1,\,(T>0)} \right|_{H=0} = -\ta{{\pi^2}/{90\,\beta^4}} + \ta{{M_{\rm eff}^2}/{24\,\beta^2}}. 
\end{eqnarray}
This is the well-known result in the high temperature expansion in flat space. 
}. 
It can be seen from \textcolor{black}{the} order-counting that Eq.(\ref{EP_03_1LZT}) diverges at the ultraviolet region. 
On the other hand Eq.(\ref{EP_03_1LFT}) does not diverge. 
%
%

We now fix $\delta m^2$ in $m_{\rm eff}^2 = m^2+ \, \delta m^2$. It can be read out from the coefficient of $\phi_0^2{/2}$.
It specifically turns out that $\delta m^2$ is given by the coefficient of $M_{\rm eff}^2$ in $V_{1,\,(T>0)}$ given \textcolor{black}{in} Eq.(\ref{EP_03_1LFT}), and the result is 
\begin{eqnarray}\label{Meff2}
m_{\rm eff}^2 =
m^2 +
\frac{\lambda}{(2\pi)^2} \int dk \,\frac{k^2}{\sqrt{9 H^2+4 k^2}}
\left(
 \frac{1}{e^{\frac{\beta}{2}  \left(\sqrt{9 H^2+4 k^2}+3 H\right)}-1} 
+\frac{1}{e^{\frac{\beta}{2}  \left(\sqrt{9 H^2+4 k^2}-3 H\right)}-1}
\right). \nonumber\\
\end{eqnarray}
\newline

We now define the effective potential in our study. For this purpose, we \textcolor{black}{think} the approximation that the space-time will be flat at $T=0$, 
although the cosmological constant in our actual cosmology is not zero. For this reason, we set our one-loop effective potential such that it can be zero 
when $T=0$ at the symmetry breaking phase. Writing such an effective potential as $ \Delta V(\beta)$, we set $\Delta V(\beta)$ as 
\begin{eqnarray}\label{Del_EP}
\Delta V(\beta) \equiv V(\beta)-C = V_{0}(\phi)-V_{0}(\phi_0) + V_{1,(T>0)},  
\end{eqnarray}
by subtracting a constant $C \equiv V_{0}(\phi_0) +V_{1,(T=0)}+\Omega$ from $V(\beta)$. Correspondingly, $Z_\phi$ is rewritten as follows:
\begin{eqnarray}\label{ToDeltaF}
Z_\phi=\exp \left( -\int_0^\beta \! d\tau \! \int d^3x \,\sqrt{g}\cdot V(\beta) \right) 
\rightarrow 
\exp \left( -\int_0^\beta \! d\tau \! \int d^3x \,\sqrt{g}\cdot \Delta V(\beta) \right). 
\end{eqnarray}
We can see that $\Omega$ given in Eq.(\ref{EP_03_1LFTC}) is diverging. In this paper we just ignore $\Omega$ in $\Delta V(\beta)$. 
\footnote{
We have performed the expansion of $M_{\rm eff}^2$ to the linear order in the above. 
Generally the high temperature expansion is an expansion with regard to the ratio of $M_{\rm eff}/T$. 
Our situation that $M_{\rm eff}$ is smaller can be regarded as the situation that $T$ is \textcolor{black}{high} 
in the ratio of $M_{\rm eff}/T$. Because of this, our expansion \textcolor{black}{can be interpreted as} the high temperature expansion.} 
\newline

We now consider the contribution from the functional integral measure. 
The functional integral measure can be considered to be written as ${\cal D}\phi =c\,\beta d\phi$, where $c$ is some dimensionless number
\footnote{
The dimension of the real scalar fields is generally $(n - 2)/2$ by counting the mass dimension, where $n$ is the space-time's dimension. 
Since $n$ is $4$ in this study, it can be seen that $[\phi]=[M^{1}]$. Integral measures in path-integrals for partition functions are 
dimensionless as ${\rm dim}[{\cal D}\phi] = {\rm dim}[{\cal M} \, d\phi] = 0$, where ${\cal D}\phi$ is functional integral measures and 
${\cal M}$ is factor parts. The dimension of the inverse temperatures is $[\beta]=[M^{-1}]$. Hence the functional integral measure can be 
written as ${\cal D}\phi = c \,\beta \,d\phi$, where $c$ is some dimensionless number. 
}. 
As a result, the contribution from the functional integral measure to the effective potential can be written in the following way
\begin{eqnarray}\label{Z_CDphi}
Z_\phi
\!\!\!&=&\!\!\! \int {\cal D} \phi \exp \left( -\int_0^\beta \! d\tau \! \int d^3x \,\sqrt{g}\cdot {\cal L}\right)\nonumber\\
\!\!\!&=&\!\!\! \int \! d\phi\, c \, \beta \exp \left( -\int_0^\beta \! d\tau \! \int \! d^3x  \,\sqrt{g}\cdot {\cal L} \right)\nonumber\\
\!\!\!&=&\!\!\! \exp \left( -\int_0^\beta \! d\tau \! \int \! d^3x \,\sqrt{g}\cdot \left(V(\beta) - \frac{1}{{\rm \ta{Vol}}}(\log \beta + \log c)\right) \right), \label{Zphi}
\end{eqnarray}
where $\displaystyle {\rm \ta{Vol}} \equiv \int_0^\beta \! d\tau \! \int \! d^3x \,\sqrt{g}$. 
We can see that the contribution from the measure does not contribute due to the infinite contribution of ${\rm \ta{Vol}}$.  
Consequently it turns out that we do not need to take into account the contribution from the functional integral measure. 
\newline

The effective potential we consider is finally Eq.(\ref{Del_EP}), and our $Z_\phi$ can be written as 
\begin{eqnarray}
Z_\phi = \exp \left( -\int_0^\beta \! d\tau \! \int d^3x \,\sqrt{g} \cdot \Delta V(\beta) \right). 
\end{eqnarray}

\section{Fixing the background space-time}
\label{Chap:Fixing the background space-time} 

In this chapter we obtain the equation to fix the background geometry by regarding the effective potential as the positive cosmological constant. 
Now our total partition function can be written as
\begin{eqnarray}\label{Our EH model in E}
Z= \exp \left( - \int_0^\beta \! d\tau \! \int \! d^3x \, \sqrt{g} \cdot \big( -R + \Delta V(\beta) \big) \right).
\end{eqnarray}

Now we give here general description for the Wick rotation to the Einstein-Hilbert case as
\begin{eqnarray}\label{General EH model in E}
iS^M 
\!\!\!&=&\!\!\! \int \! d^4x \, \sqrt{-g^M} \, \left( R^M-2\Lambda^M \right)\nonumber\\ 
\!\!\!&\rightarrow&\!\!\!
  \int \! d\tau \! \int \! d^3x \, \sqrt{g} \,  \left(R - 2\Lambda \right)\nonumber\\ 
\!\!\!&=&\!\!\! - \int \! d\tau \! \int \! d^3x \, \sqrt{g} \, \left(-R + 2\Lambda \right) = -S\ta{.}
\end{eqnarray}

Backing to our story, we can confirm that the Einstein equation in the Euclid de Sitter space (\ref{Euclid de sitter}) can be satisfied when $\Lambda=-3H^2$. 
Hence the $-R + 2\Lambda$ in the last line of Eq.(\ref{General EH model in E}) can be written as $-R - 6H^2$ in our study. 
As a result, comparing Eqs.(\ref{Our EH model in E}) and (\ref{General EH model in E}), we can have the following equation: 
\begin{eqnarray}\label{CosmoConstfromDeltaF}
6H^2 = - \Delta V(\beta) \big|_{\phi=\phi_0}. 
\end{eqnarray}
We can also have teh equation:~$\partial\left(\Delta V(\beta)\right)/\partial \phi=0$ to determine the condensation $\phi_0$ at the spontaneous symmetry breaking phase. 

Now we have two unknown variables, $H$ and $\phi_0$, and we have two equations. 

\section{Space-time, and energy corresponding to the change of the acceleration and particle's world-line}
\label{Chap:Result} 

\textcolor{black}{We solve Eq.(\ref{CosmoConstfromDeltaF}) and $\partial\left(\Delta V(\beta)\right)/\partial \phi=0$ simultaneously by the numerical way.  
As a result, we obtain the results of the (total) effective potential $R-\Delta V$ and the Hubble constant $H$ with the contributions of each other, against the temperature.} 
Based on these results, we perform an evaluation of the accelerating particle's world-line and the amount of the energy to \textcolor{black}{grow} 
the temperature at each temperature. Here the temperature in our results is thought as the Unruh temperature (\ref{TUnruh}).

In our \textcolor{black}{results}, the red results represent the results obtained by solving for the two variables $H$ and $\phi_0$. 
On the other hand, the blue results represent the results obtained by solving only for $\phi_0$ with fixing $H$ to zero. 
Hence the blue points are the results in the flat space.

The parameters ($m^2$,\,$\lambda$) are always taken to be ($-1$,\,$1$). At this time, it turns out that the critical temperature $T_c$ 
for the symmetry breaking restoration is given as $T_c \simeq 4.402482673$. Here, we determined $T_c$ from whether the values of 
the squared effective mass $m_{\rm eff}^2$ (\ref{Meff2}) and the condensation $\phi$ are numerically zero or not.

\subsection{Effective potential and space-time}
\label{Chap:Result01} 

\begin{figure}[!h]
\begin{center}
\includegraphics[width=79mm,clip]{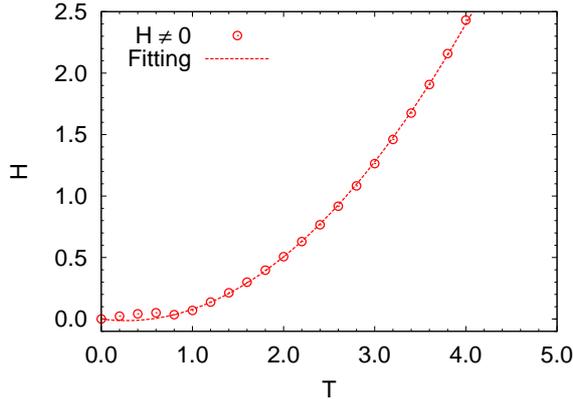}
\end{center}
\caption{The Hubble constant $H$ defined in Eq.(\ref{de sitter}) against the temperature $T$. The dotted line is the fitting result.}
\label{Fig_H} 
\end{figure}

Let us first consider $H$ against $T$. The points in Fig.\ref{Fig_H} are the calculated results for $H$, 
and the dotted line is a fitting result. The fitting function we have adopted is $c_1\,T + c_2\,T^2$, where $c_1$ and $c_2$ are the fitting parameters.
These are fixed as $(c_1,c_2)=(-0.0943597,0.173624)$ in our fitting.    

We now have obtained the Hubble constant. We can see that, as $T$ gets higher, $H$ grows as in Fig.\ref{Fig_H}. 
Hence we can see that, as the temperature gets higher, the background space transforms to the de Sitter space gradually.

\begin{figure}[!h]
\begin{center}
\includegraphics[width=80mm,clip]{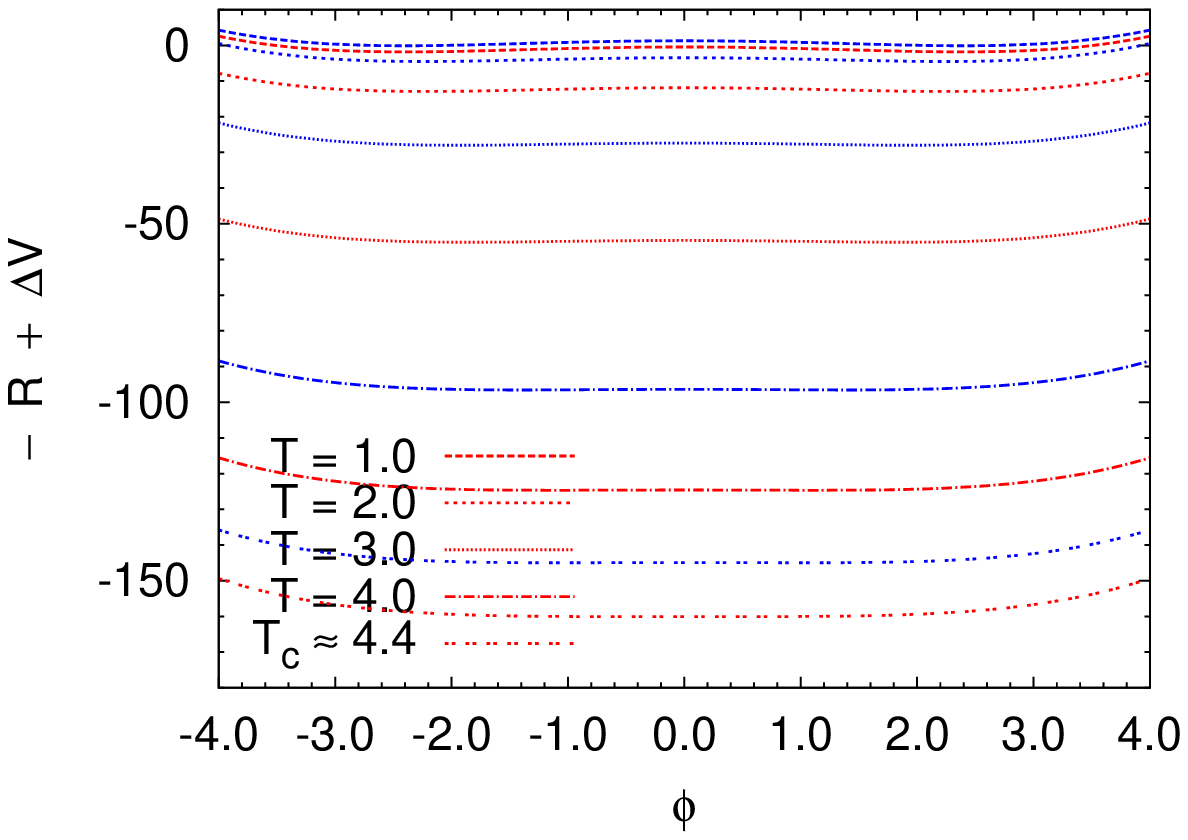} 
\includegraphics[width=79mm,clip]{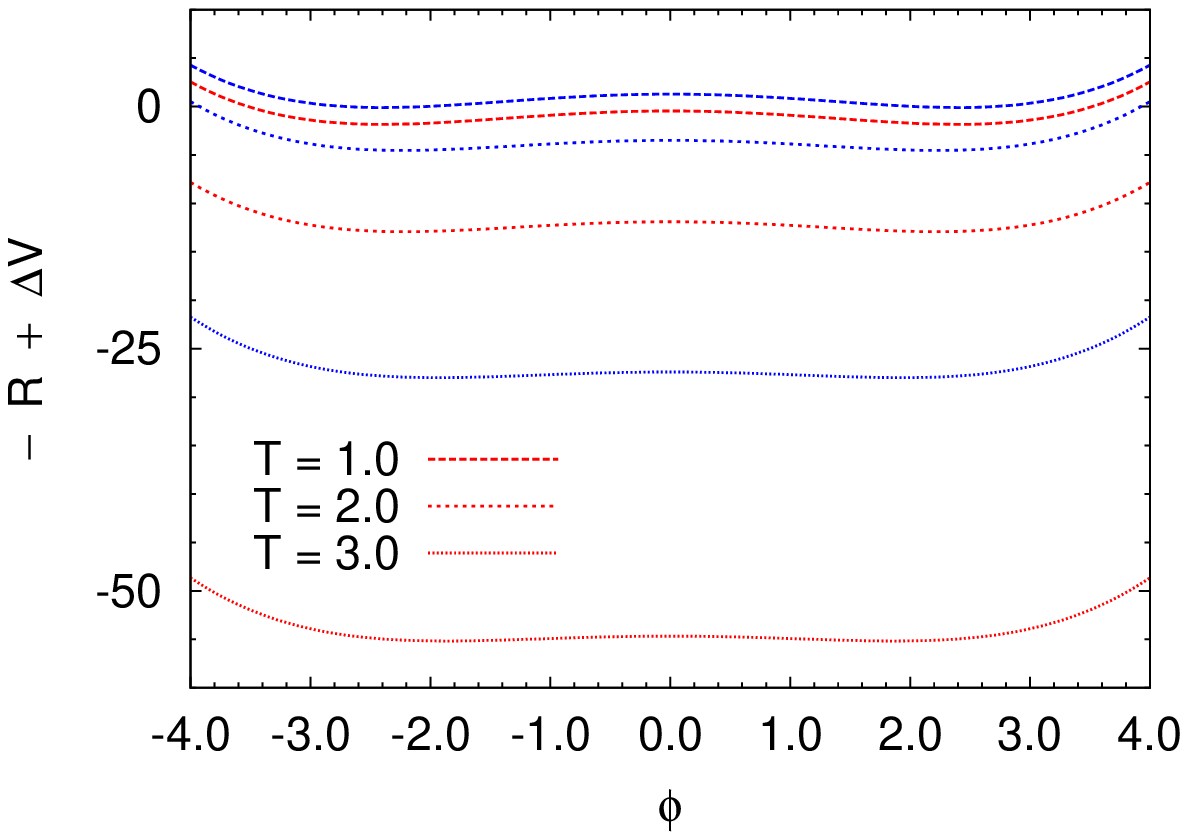} 
\includegraphics[width=79mm,clip]{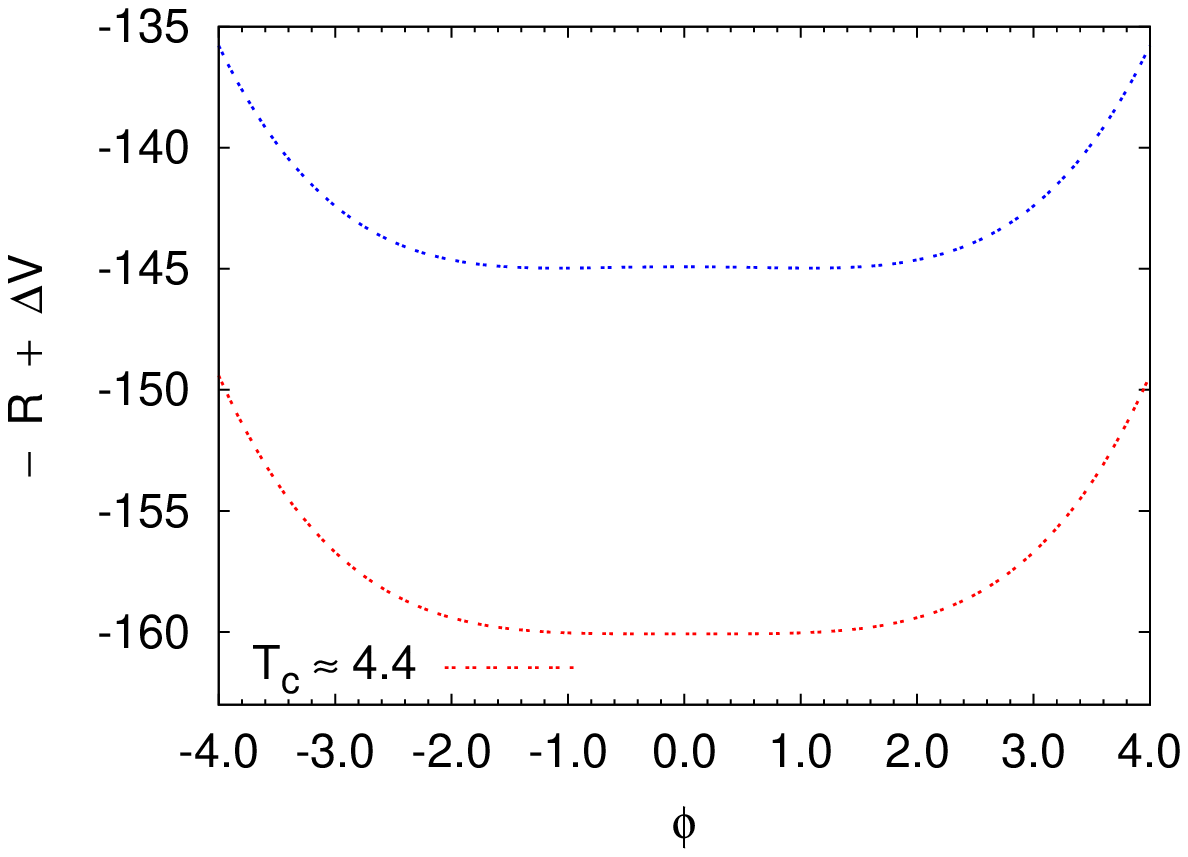} 
\end{center}
\caption{
The effective potential $\Delta V$ defined in Eq.(\ref{Del_EP}) against the condensation value $\phi$ at various $T$. 
The red lines represent the results with the background space's contribution ($H\not=0$), 
and the blue lines represent the results without the background space's contribution ($H=0$). 
Although there is only an explanation for the red lines in the caption of the figure, as to the blue lines, 
the same kind of line is used for the results calculated at the same temperatures. 
The \textcolor{black}{bottom two figures are} just enlarged views of the entire figure in the top so that the dense and faint results can be shown well.
}
\label{Fig_DV}
\end{figure}

\begin{figure}[!h] 
\begin{center}
\includegraphics[width=79mm,clip]{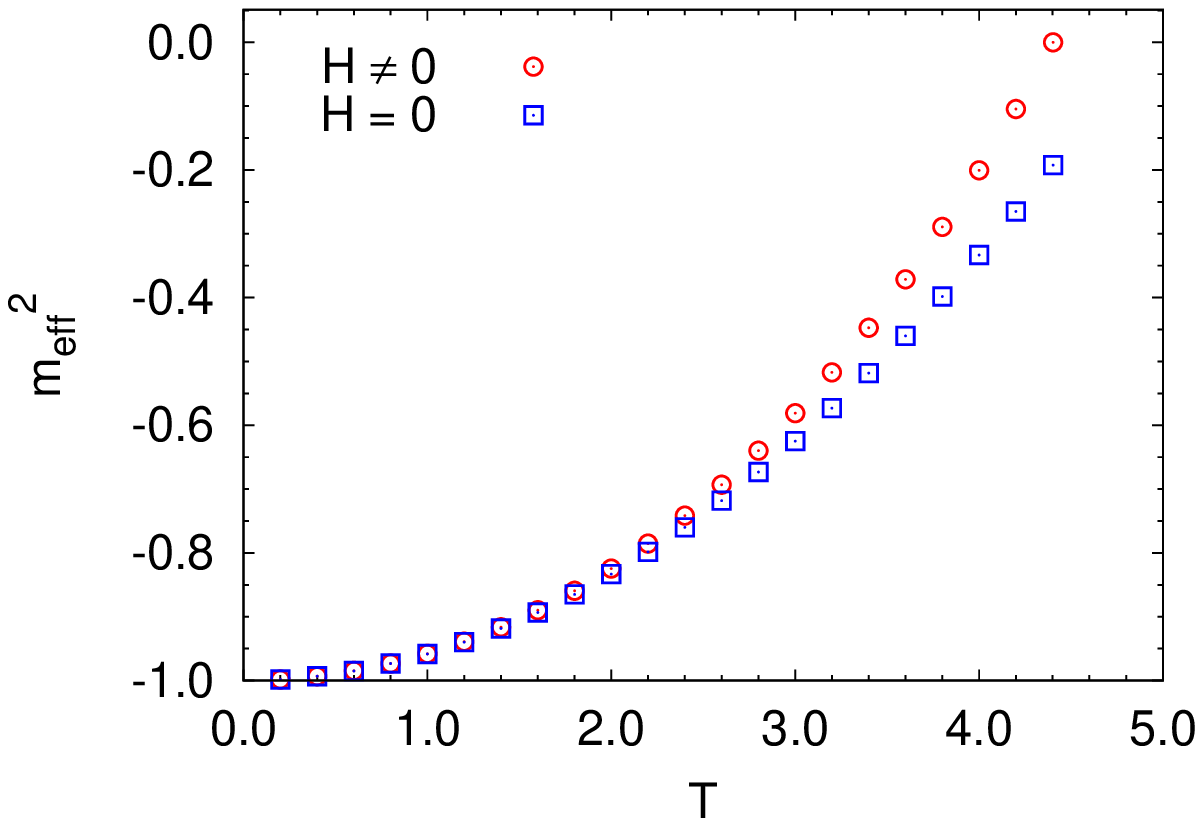}
\includegraphics[width=79mm,clip]{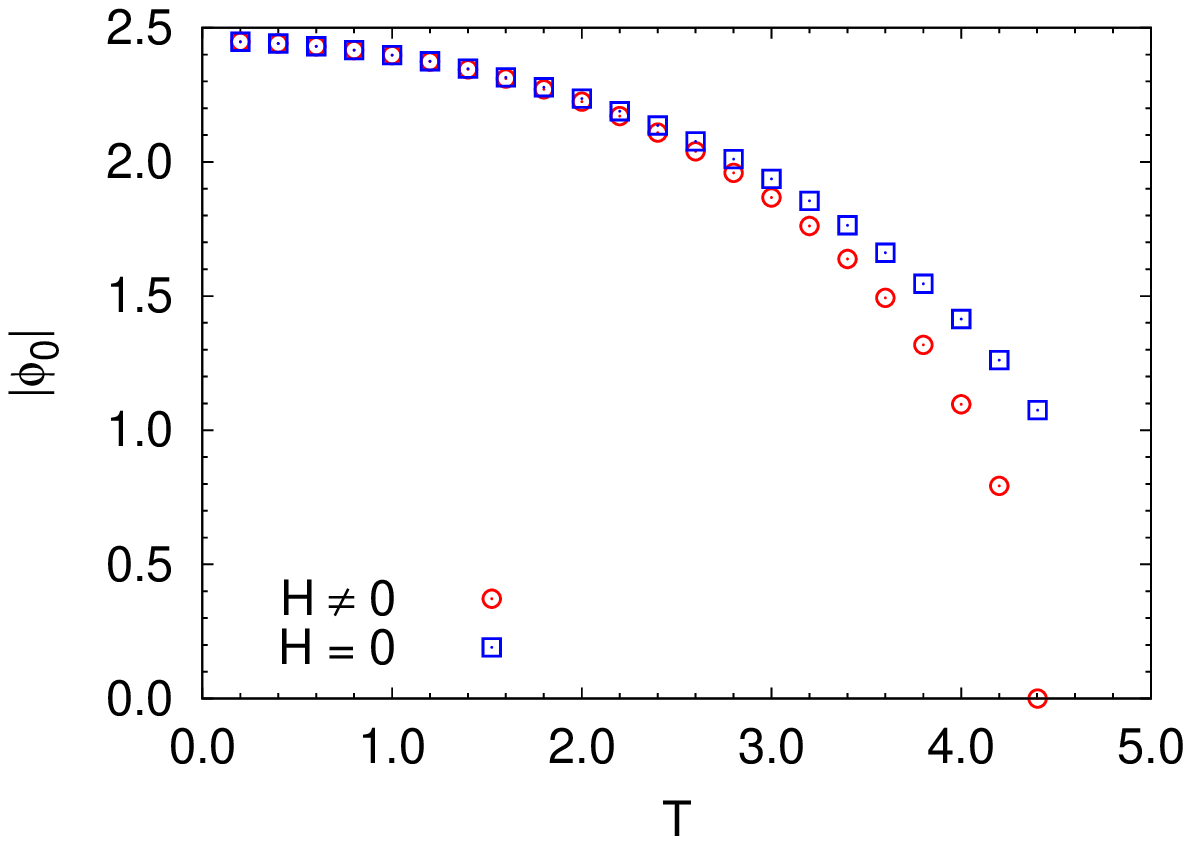} 
\end{center}
\caption{The squared effective mass $m_{\rm eff}^2$ in Eq.(\ref{Meff2}) and the condensation $|\phi_0|$ against the temperature $T$.}
\label{Fig_M2eff}
\end{figure}

\textcolor{black}{Next, in Fig.\ref{Fig_DV} the (total) effective potential $R-\Delta V$, 
where the scalar curvature $R$ in the background space in Eq.(\ref{Euclid de sitter}) is given by $-12H^2$.} 
We can see from Fig.\ref{Fig_DV} that the background space's effect lowers the critical temperature. We show Fig.\ref{Fig_M2eff}~
(the numerical results of the squared effective mass $m_{\rm eff}^2$ and the field's condensation $|\phi_0|$) to support this. 
\footnote{\textcolor{black}{We should notice that we have performed the expansion equivalent to the high temperature expansion as mentioned under Eq.(\ref{Del_EP}). 
Hence our analysis gets more better 
in the higher temperature region that satisfies $|M_{\rm eff}/T| \ll 1$ .}}

%
%

%
%
%
%
%
%
%
%
%
%
%
%

\subsection{Energy corresponding to the change of the acceleration and particle's world-line}
\label{Chap:Result02} 

\begin{figure}[!h] 
\begin{center}
\includegraphics[width=79mm,clip]{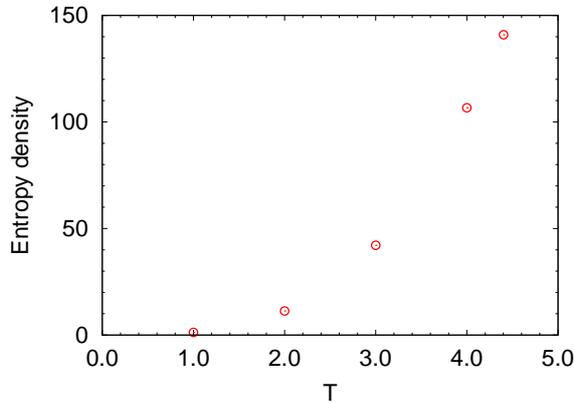}
\end{center}
\caption{The entropy density against the temperature  $T$, obtained by calculating $\partial F /\partial T$ numerically, where $F=-R +\Delta V$ in Fig.\ref{Fig_DV}.}
\label{Fig_Entropy}
\end{figure}

The effective potential we have calculated can be regarded as a free energy. We can hence read out the entropy density according to the general relation 
$S = -\partial F / \partial T$~($S$ and $F$ are general entropy densities and free energies, \textcolor{black}{respectively.}). 
We actually have evaluated the entropy density numerically at several temperatures. These results are shown in Fig.\ref{Fig_Entropy}.

\textcolor{black}{
The entropy density in Fig.\ref{Fig_Entropy} can be interpreted as the amount of the energy for the growing of the unit temperature at each temperature. 
We further consider in this study that the temperature of the particle in the inertial frame that is rooted in the Unruh effect is to be simply proportional 
to the acceleration through the Unruh effect~(\ref{TUnruh}). Hence we can consider that the amount of energy for the growing of the unit acceleration at each 
acceleration is proportional to the entropy density in Fig.\ref{Fig_Entropy} at each temperature to which the acceleration corresponds. As a result we can see 
from Fig.\ref{Fig_Entropy} that the higher the acceleration gets, the more energy is required.}
%
%
%
%
%
%
\newline

\begin{figure}[!h]
\begin{center}
\includegraphics[width=79mm,clip]{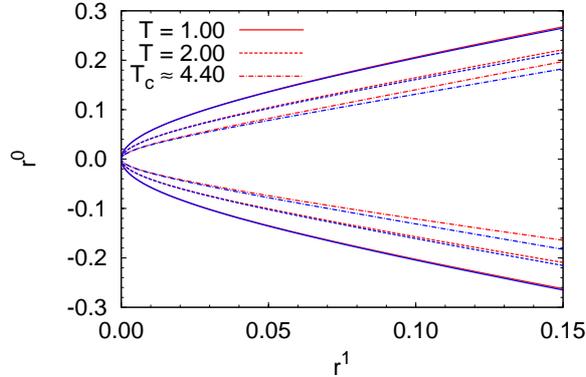}
\end{center}
\caption{
The world-lines of the uniformly accelerating particle at various $T$. 
The red lines represent the results with the background space's contribution~($H\not=0$), and the blue lines represent the results 
without the background space's contribution~($H=0$). Although there is only an explanation for the red lines in the caption of the figure, 
for the blue lines, the same kinds of line are used for the results calculated at the same temperatures as well as Fig.\ref{Fig_DV}.
}
\label{Fig_Orbit}
\end{figure}

We next turn to the issue of the accelerating particle's world-line. In our study, the space-time in the neighborhood of the accelerating particle is predicted 
to be the de Sitter space \textcolor{black}{in Eq.(\ref{de sitter}),  which is characterized by $H$ and its behavior is given in Fig.\ref{Fig_H}.}  
Hence it is assumed that the accelerating particle's world-line gets a correction compared with the case of a flat space. To investigate \textcolor{black}{these}, writing the position of 
the particle as $r^\mu(\tau)$~($\tau$ is the proper time), we solve the geodesic equation in the de Sitter space-time (\ref{de sitter}) with a force added for the uniform acceleration of 
the particle, 
$F^\mu = ma\,({r^1}'(\tau),{r^0}'(\tau),0,0)$~\textcolor{black}{(This is the same way with the one in Ref.\cite{Iso:2010yq}.)}. Concretely, 
\begin{eqnarray}
e^{2H\tau}H \, ({r^1}')^2+{r^0}'' \!\!\!&=&\!\!\! F^0,\\
2H \, {r^0}' + {r^1}'' \!\!\!&=&\!\!\! F^1,
\end{eqnarray}
where we have confined the motion of the particle to the $(0,1)$-plane.  
$H$ is assumed to have the temperature dependence obtained from the fitting in the beginning of Sect.\ref{Chap:Result01}, 
and $a$ is proportional to $T$ through the Unruh effect (\ref{TUnruh}).   
We solve the above equations of motion numerically with the boundary condition:
\begin{eqnarray}
\big(r^0(0),{r^0}'(0)\big) = (0,1) \quad {\rm and} \quad \big(r^1(0), {r^1}'(0)\big) =(0,0) 
\end{eqnarray} 
at several $T$. 
We show its results in Fig.\ref{Fig_Orbit}. 

\textcolor{black}{In Fig.\ref{Fig_Orbit}, the horizontal and longitudinal axes mean the direction of the motion and the time respectively. In the figure we can see how the accelerations 
to decrease and increase the speed of the particle are from how amount of the speed~(that can be read out from the gradient of the line) can be vanished in a same distance toward the 
coming particle and how long time is taken to move a same distance toward the separating particle, respectively. Here, the velocity of the particle reaches zero at $r^1=0$. From these 
consideration, we can see from the figure that the Unruh thermal radiation has the effect to grow the coming particle's acceleration~(the acceleration to decrease the speed) while diminish 
the separating particle's acceleration~(the acceleration to increase the speed).} 
%
%
%
%

\section{Remark}
\label{Chap:Remark} 

In this paper, corresponding to the prediction in Refs.\cite{Experiment,Sciama,Raval:1995mb,Chen:1998kp,Iso:2010yq} 
that an accelerating particle performs a Brownian motion rooted in the Unruh effect in the inertial frame,  
we have \textcolor{black}{proposed a prediction} that, although the Unruh effect itself occurs only in the accelerating frame, 
the heat taken in the body of the particle in the accelerating frame by the Unruh effect could survive even in the inertial frame, 
and the resultant thermal radiation could appear from an accelerating particle in the inertial frame. 
We have named such a radiation ``{\it Unruh thermal radiation}'' in the fourth paragraph of Sect.\ref{Chap:Intro}.

We then have calculated the one-loop effective potential at finite temperature taking into account that the background space-time is deformed from the flat space-time 
to the de Sitter space-time due to the field theory's corrected one-loop effective potential. In this paper, \textcolor{black}{considering that the temperature of the 
accelerating particle in the inertial frame would be proportional to the Unruh temperature}, we have simply \textcolor{black}{thought} the temperature of the accelerating particle 
as the Unruh temperature. The model we have considered has been a real scalar field described by the four-dimensional Klein-Gordon model with a \textcolor{black}{Higgs type} potential term 
at finite temperature.

As for technical \textcolor{black}{staffs}, we now touch on a problem to be noticed particularly. 
It is the region where the Unruh thermal radiation can reach and the temperature distribution in that region. 
We have taken the radial direction in that region from zero to infinity, and regarded the temperature distribution to be uniform in the entire region. 
However, in a realistic world, it should be finite and the temperature gradually decline \textcolor{black}{as getting away} from the particle~(working as the thermal source).  
But if these were taken into account in the analysis, it is expected that the factoring out of the volume factor done in Eq.(\ref{VolVE}) becomes a problem. 
As a result, the analysis would become highly complicated. 
We are going to treat these issues more precisely in our future work.

As for our results, when a deformed background space-time is taken, it turns out that the critical temperature becomes smaller than the case that the background space-time 
is fixed to the static flat case. We have further revealed that the amount of the energy \textcolor{black}{to grow the unit acceleration at each acceleration increases 
as the acceleration gets higher, and how the correction is in the world-line of the accelerating particle}.

It would be intriguing to examine how our results turn out in the future as we develop the model to more realistic cases,
and to confirm whether the tendencies obtained in our study are consistent or not compared with the experiments. 
Furthermore, the Unruh effect has a close relation with the Hawking temperature. 
The Hawking temperature at the moment is a result of the semi-classical gravity, \textcolor{black}{
and currently we could not exclude the possibility that it gets some correction in the full quantum gravity.}  
In this sense, the future development of this study would also be intriguing.

\section*{Acknowledgment}
S.T. wishes to thank Burin Gumjudpai, \textcolor{black}{Kazuhiro Yamamoto}, Pichet Vanichchapongjaroen, Seckson Sukkhasena, \textcolor{black}{Sen Zhang}, 
Shinji Tsujikawa, and Shouichi Ichinose for helpful discussions, advice, and various things. S.T. further 
would like to offer thanks to the Institute for Fundamental Study and Naresuan University. The numerical 
calculations in this paper were performed in the cluster system at the high energy theory group of National Technical 
University of Athena managed by Konstantinos Anagnostopoulos.


\end{document}